\documentclass[twocolumn,showpacs,preprintnumbers,prl,aps,amssymb]{revtex4}
\usepackage{graphicx}% Include figure files
\usepackage{dcolumn}% Align table columns on decimal point
\usepackage{bm}% bold math
\usepackage{stmaryrd}
\begin{document}

%\preprint{preprint}

%%%%%%%%%%%%%%%%%%%%%%%%%%%% TITLE

\title{Field-Induced Quantum Critical Point in CeCoIn$_5$}

%%%%%%%%%%%%%%%%%%%%%%%%%%%% AUTHORS
\author{Johnpierre~Paglione}
\email{paglione@physics.utoronto.ca}
\author{M.~A.~Tanatar}
\altaffiliation[Permanent address: ] {Inst. Surface Chemistry, N.A.S. Ukraine, Kyiv, Ukraine.}
\author{D.~G.~Hawthorn, Etienne~Boaknin, R.~W.~Hill, F.~Ronning, M.~Sutherland}
\author{Louis Taillefer}
\altaffiliation[Current address: ] {D\'epartement de physique, Universit\'e de Sherbrooke,
Sherbrooke, Canada, J1K 2R1.}
\affiliation{Department of Physics, University of Toronto, Toronto, Ontario M5S 1A7, Canada}
\author{C.~Petrovic}
\altaffiliation[Current address: ] {Department of Physics, Brookhaven National Laboratory, Upton,
New York, 11973.}
\author{P.~C.~Canfield}
\affiliation{Ames Laboratory and Department of Physics and Astronomy, Iowa State University,
Ames, Iowa, 50011}

\date{\today}

%%%%%%%%%%%%%%%%%%%%%%%%%%%% ABSTRACT

\begin{abstract}
The resistivity of the heavy-fermion superconductor CeCoIn$_5$ was measured as a function of
temperature, down to 25 mK and in magnetic fields of up to 16 T applied perpendicular to the
basal plane. With increasing field, we observe a suppression of the non-Fermi liquid behavior,
$\rho \sim T$, and the development of a Fermi liquid state, with its characteristic
$\rho=\rho_0+AT^2$ dependence. The field dependence of the $T^2$ coefficient shows critical
behavior with an exponent of 1.37. This is evidence for a field-induced quantum critical point
(QCP), occuring at a critical field which coincides, within experimental accuracy, with the
superconducting critical field $H_{c2}$. We discuss the relation of this field-tuned QCP to a
change in the magnetic state, seen as a change in magnetoresistance from positive to negative, at
a crossover line that has a common border with the superconducting region below $\sim 1$~K.
\end{abstract}

\pacs{71.27.+a,73.43.Qt,75.40.-s}
\maketitle

%%%%%%%%%%%%%%%%%%%%%%%%%%%% INTRODUCTION

The recent discovery of a new family of heavy-fermion superconductors with the general formula
CeMIn$_5$ (M=Co, Ir, Rh or their solid solutions) has generated much interest. The ground state of
these compounds can be fine-tuned between magnetic order, superconductivity, and a coexistence of
the two by varying both pressure and alloy composition \cite{alloy}. CeCoIn$_5$, with the highest
ambient pressure superconducting transition temperature $T_c$ of all heavy-fermion compounds
discovered to date \cite{Cedomir}, is believed to be close to a point where the magnetic state
becomes unstable as $T\rightarrow 0$ \cite{Nicklas,Shishido,Sidorov}. This proximity to a quantum
critical point (QCP) is believed to be responsible for the unusual properties observed in
the superconducting \cite{Movshovich,Kohori,Curro,Izawa,penetration} and
normal \cite{Sarrao-review} states.

The peculiar magnetic properties of CeCoIn$_5$ are determined by the magnetic moments of
Ce$^{3+}$ ions and by conduction electrons. Through systematic studies of
Ce$_{1-x}$La$_x$CoIn$_5$ alloys, where the dilution of magnetic Ce ions by La (non-magnetic
analog) allows one to study the nature of intersite interactions, it has been shown that the
energy scales associated with the relevant magnetic interactions are all well separated, and that
the dominance of direct intersite interactions below the coherence peak temperature $T_{coh}
\sim~$50~K gives rise to pronounced two-dimensional antiferromagnetic (AF) correlations
\cite{Nakatsuji}. Although long-range magnetic order is not present in CeCoIn$_5$ \cite{uSR}, the
close proximity of this system to AF order \cite{Kohori} results in an abundance of spin
fluctuations which lead to behavior that is notably different from that expected in the Fermi
liquid (FL) model.

In addition to the observation of $T-$linear resistivity \cite{Cedomir} commonly associated
with quantum criticality, non-Fermi liquid (NFL) behavior manifests itself in CeCoIn$_5$ in a
number of ways. These include a logarithmic increase of the electronic specific heat
coefficient on cooling \cite{specificheat}, an enhancement of the effective mass at low
temperatures and its strong field dependence, as seen in de~Haas-van~Alphen \cite{dHvA} and
microwave conductivity \cite{Ozcan} experiments, and a magnetic susceptibility that does not
saturate at low temperatures \cite{Cedomir}.

The close proximity of CeCoIn$_5$ to an AF QCP at ambient pressure offers a unique opportunity to
explore quantum critical phenomena in a system free of disorder due to alloying. Since magnetic
fluctuations play an essential role in quantum criticality, the response of this system to
applied magnetic fields is of clear interest. We have performed a systematic study of the low
temperature electrical resistivity of CeCoIn$_5$ in magnetic fields up to 16~T. We find that this
system can indeed be driven through a QCP by a magnetic field, as evidenced by a divergence of the
electron-electron scattering strength at the critical field and the subsequent restoration of a FL
state. We show that magnetic interactions appear to play a key role in this critical behavior.

%%%%%%%%%%%%%%%%%%%%%%%%%%%% EXPERIMENTAL

Single crystals of CeCoIn$_5$ were grown by the self-flux method \cite{Cedomir}. As-grown
crystals have a thin platelet shape, with large surfaces corresponding to the (001) basal plane.
For this study we have cut four samples into rectangular parallelepipeds with typical dimensions
$\sim 2 \times 1 \times 0.3$~mm. Electrical contacts for standard four-wire measurements were
made with soldered indium, resulting in $\sim 5$~m$\Omega$ contact resistances. The in-plane
electrical resistivity, $\rho$, was measured with an AC resistance bridge by applying $0.1$~mA
excitation currents along the basal plane, and transverse magnetic fields up to 16~T parallel to
the [001] axis. Measurements were performed from 300~K to 0.3~K in a $^3$He cryostat, and down to
25~mK in a $^3$He/$^4$He dilution refrigerator. The 16~T data were reproduced in reversed field
direction, excluding any spurious Hall component.

%%%%%%%%%%%%%%%%%%%%%%%%%%%% RESULTS AND DISCUSSION

\begin{figure}
 \centering
 \includegraphics[totalheight=3.8in]{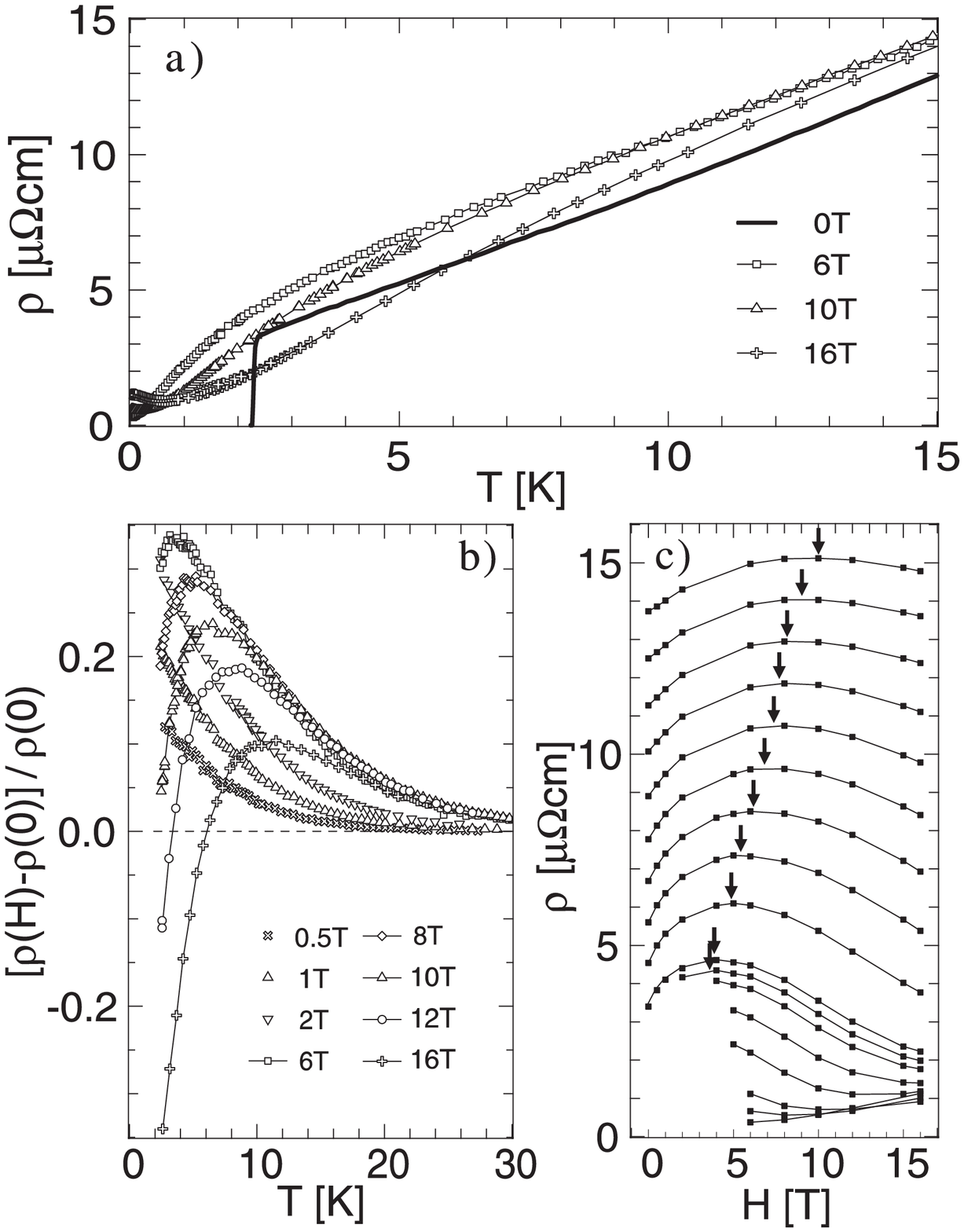}
 \caption{\label{fig:highT} Resistivity of CeCoIn$_5$ under various fields, shown a) for $H=$~0, 6,
 and 16~T, b) as magnetoresistance versus temperature, normalized to zero field resistivity, and c)
 versus applied field, as isotherms for $T=$ 16, 14.5, 13, 11.5, 10, 8.5, 7, 5.5, 4, 2.5, 2.25, 2,
 1.5, 1, 0.5, 0.3 and 0.1~K from top to bottom (lines are guides to the eye and are not offset).
 The arrows in c) indicate the position of the crossover from positive to negative MR with
 increasing field for each temperature.}
\end{figure}

On cooling from room temperature, the zero-field resistivity of CeCoIn$_5$ shows a slight
increase, followed by a crossover to metallic behavior below $T_{coh}$, as is seen in many
heavy-fermion systems \cite{Stewart}. Below $\sim 10$~K, $\rho(T)$ displays a linear
temperature dependence down to $T_c=2.3$~K, as shown in Fig.~\ref{fig:highT}a. Such $T$-linear
behavior has been observed in many systems which lie at or close to a QCP \cite{Stewart,QCP}.
Above 30~K, the magnetoresistance (MR) is negligibly small up to 16~T, whereas below 30~K a
notable MR begins to develop as shown in Fig.~\ref{fig:highT}b. The linear dependence of
$\rho(T)$ observed at $H=0$ suffers a drastic change with increasing field, most pronounced at
low temperatures. For instance, the 6~T curve (Fig.~\ref{fig:highT}a) displays a notable
deviation from linearity below $\sim$ 5~K and decreases on cooling to a value much lower than
the residual resistivity inferred from a linear extrapolation of $\rho(T,H$=0) to $T$=0. By
16~T, the downturn in $\rho(T)$ shifts to much higher temperatures (Fig.~\ref{fig:highT}a), as
evidenced by the large negative MR below $\sim$ 5~K shown in Fig.~\ref{fig:highT}b. At higher
temperatures, the field dependence of $\rho$, plotted at constant $T$ values in Fig.
\ref{fig:highT}c, reveals the development of a {\it crossover} in the sign of MR with
increasing field and temperature, which will be discussed below.

\begin{figure}
 \centering
 \includegraphics[totalheight=2.6in]{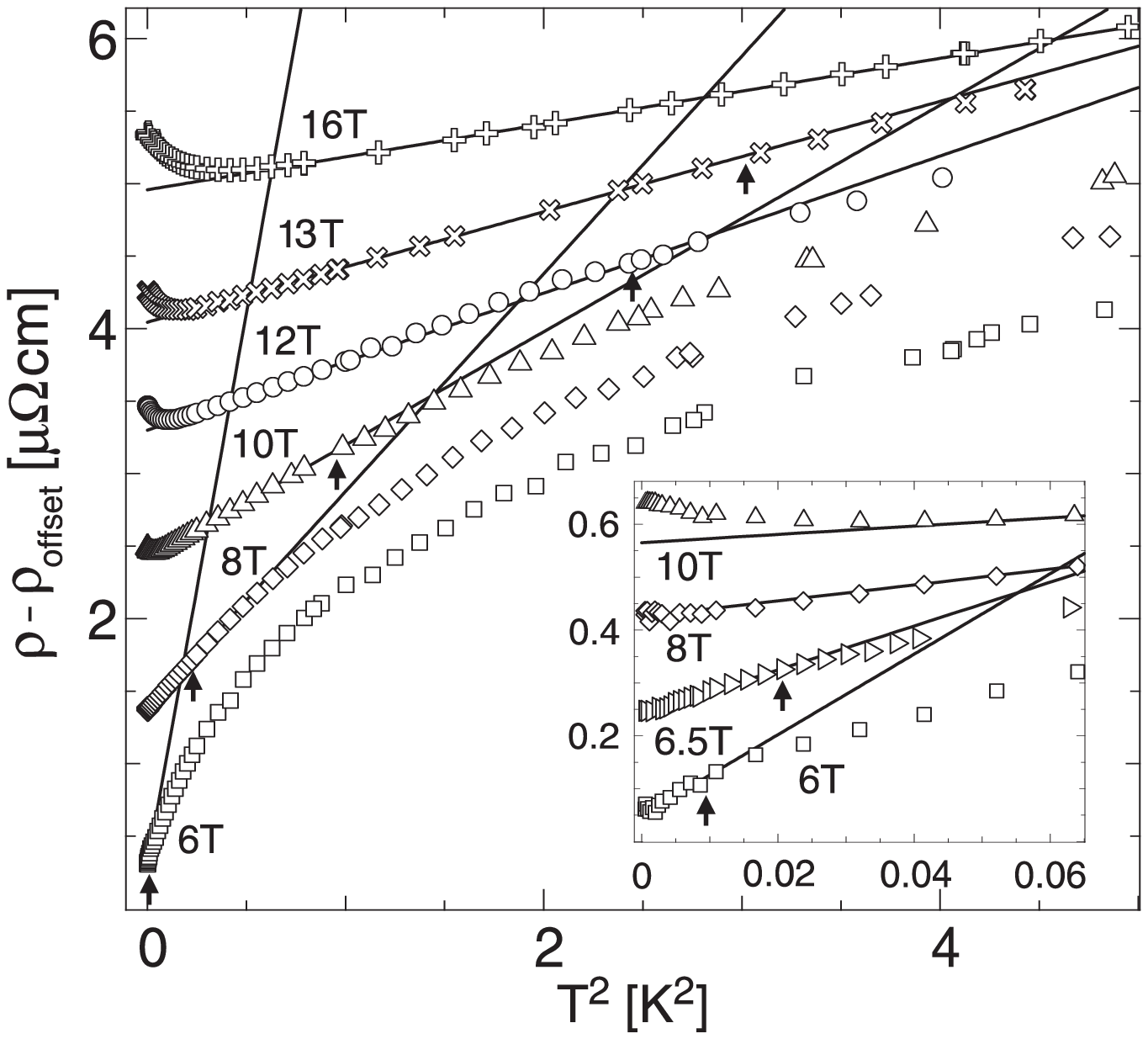}
 \caption{\label{fig:lowT} Low temperature resistivity of CeCoIn$_5$ plotted vs. $T^2$ for several
  magnetic fields, with an inset showing low temperature data (data sets are offset for clarity in
  both figures). The solid lines are linear fits to the data, showing the behavior of the
  quadratic-term coefficient (slope) under applied field, with arrows indicating the upper limit of
  the temperature range of $T^2$ behavior.}
\end{figure}

A close analysis of our 6~T data at low temperatures reveals a narrow but clearly distinguishable
range of $T^2$ behavior below $\sim$100~mK, highlighted in a plot of $\rho$ vs. $T^2$ in Fig.~\ref{
fig:lowT}. This range gradually becomes wider and more apparent with increasing field, as shown by
the linear fits \cite{fits} in the main panel of Fig.~\ref{fig:lowT}, and extends to as high as 2.5
~K by $H$=16~T. Simultaneously, at the lowest measured temperatures, a small {\it upturn} in
$\rho(T)$ starts to develop above 8~T and continues to grow upon further field increase. Since this
effect is confined to very low temperatures and high fields, it does not hinder the observation of
$T^2$ resistivity and is left for future study \cite{upturn}.

%%%%%%%%%%%%% figure 1abc

The behavior of MR in CeCoIn$_5$ can be compared to that observed in two closely related
superconductors. In CeIrIn$_5$, no significant MR has been observed between 50~mK and 5~K at
ambient pressure \cite{Cedomir}. In CeRhIn$_5$, a large positive MR was observed
\cite{Christianson} at ambient pressure, while under the critical pressure (where AF order is
suppressed and superconductivity is dominant) the low temperature MR \cite{Muramatsu} bears a
striking similarity to that observed in our study. The development of a $T^2$ dependence of
resistivity was also seen in CeCoIn$_5$ under applied pressure, where a jump in the exponent of
$T$ from linear to quadratic occurs near 2~GPa \cite{Sidorov}. From these comparisons, it would
appear that the Co system at ambient pressure is close to quantum criticality in the same manner
as the Rh system under critical pressure, while the Ir system is further from criticality at
ambient pressure.

%%%%%%%%%%%%% figure 2

\begin{figure}
 \centering
 \includegraphics[totalheight=2.7in]{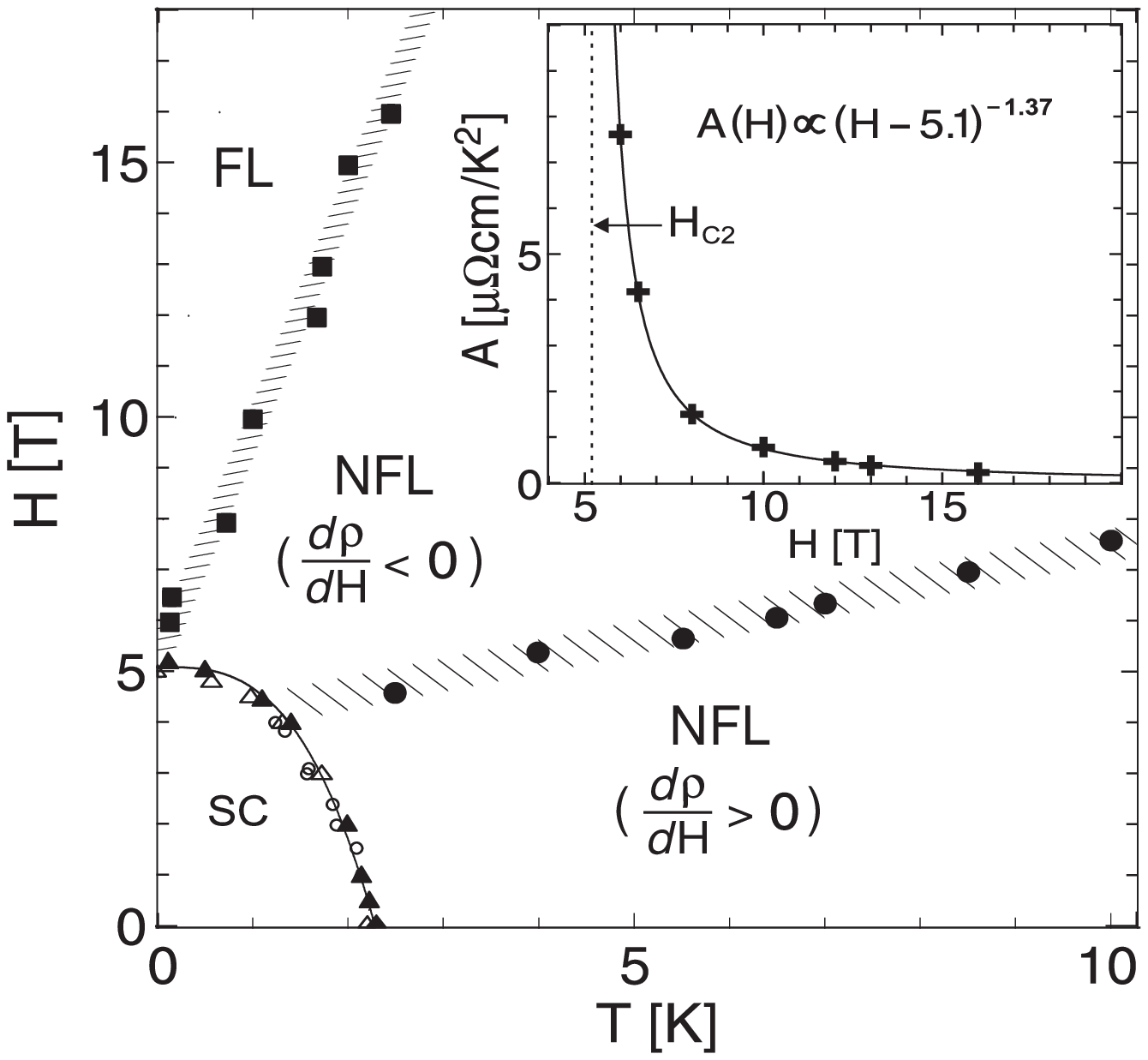}
 \caption{\label{fig:phase} $H-T$ phase diagram of CeCoIn$_5$ determined
from resistivity measurements, including the upper bound of $T^2$ resistivity
($\blacksquare$), the position of the MR maximum ($\bullet$), and $H_{c2}$ determined by
resistivity $H$- and $T$-sweeps ($\blacktriangle$), which agrees with specific heat ($\circ$)
and resistivity ($\vartriangle$) measurements from Ref.~\onlinecite{Cedomir}. The hatched
lines indicate the crossover boundaries between FL and NFL (negative and positive MR) regimes,
as explained in the text. The inset shows the field dependence of the quadratic coefficient
$A$ of $\rho(T)$ (solid line is a fit of the data points ($\bf +$) to the displayed formula).}
\end{figure}

The slope of the fitted $\rho$ vs. $T^2$ curves (Fig.\ref{fig:lowT}), {\it i.e.} the
coefficient $A$ of the $T^2$ term (in $\rho = \rho_0 + A T^2$), is a measure of the strength
of electron-electron interactions, notoriously high in heavy-fermion materials. As is clear
from the fits in Fig.~\ref{fig:lowT}, $A$ tends to decrease with increasing field. The field
dependence of $A$, or $A(H)$ (inset in Fig.~\ref{fig:phase}), displays {\it critical} behavior
best fitted by the function $A\propto (H-H^*)^{\alpha}$, with parameters $H^*=5.1\pm0.2$~T and
$\alpha=-1.37\pm0.1$. In FL theory, the coefficient $A$ is roughly proportional to the square
of the electronic contribution to the specific heat coefficient $\gamma$. Thus, the critical
behavior of $A(H)$ would lead us to expect a divergence of $\gamma(H)$ somewhere close to
$H_{c2}$. Experimental studies of the specific heat at fields slightly exceeding $H_{c2}$
indeed show an increase of $\gamma(H)$ \cite{specificheat}, but unfortunately detailed data is
lacking in the field range of interest to allow a direct comparison.

The field-induced recovery of a FL regime in CeCoIn$_5$ exhibits a distinct similarity to the
behavior observed in several other systems. In Sr$_3$Ru$_2$O$_7$ and CeRu$_2$Si$_2$, a
field-induced anomaly in resistivity is associated in both cases with a change from
predominantly AF to ferromagnetic fluctuations \cite{Flouquet}. In
U$_{0.9}$Th$_{0.1}$Be$_{13}$ \cite{Dickey}, a system close to a superconducting phase, the
evolution of $A(H)$ and $T^2$ resistivity with field both bear a close resemblance to that
found in our study. In YbRh$_2$Si$_2$, critical behavior in $A(H)$ was observed in proximity
to a field-induced QCP associated with a second-order AF transition, with an exponent $\alpha=-1$
extracted from $A(H)$ that is similar to the value we obtain in CeCoIn$_5$ \cite{Gegenwart}.

All of the aforementioned systems exhibit critical behavior in resistivity when approaching
some critical field value $H^*$. However, what is unique (and intriguing) about CeCoIn$_5$ is
the fact that $H^*$ is very close to $H_{c2}(0)$, which points to the existence of a QCP
coincident with the superconducting transition at $T=0$. The question is whether this
coincidence is essential or accidental. In the latter case, the critical behavior would
originate from proximity to an ordered phase (and transition) other than the superconducting
state (and in fact would be masked by superconductivity). In this respect it is interesting to
mention the compelling evidence from magnetization \cite{Murphy,Tayama}, specific heat
\cite{firstorder}, and thermal conductivity \cite{Izawa} that the $H_{c2}$ transition in
CeCoIn$_5$ is first-order below $\sim 0.7$~K for all field orientations. This suggests that
critical behavior of the kind observed here, which is usually associated with a second-order
phase transition brought to absolute zero, is not caused by the vicinity of the
superconducting state itself. Rather, it is tempting to propose that the quantum critical
behavior observed in CeCoIn$_5$ is associated with a zero-temperature transition of magnetic
origin, much as the field-induced transition from an AF state to a field-aligned state in
YbRh$_2$Si$_2$.

%%%%%%%%%%%%% figure 3

In connection to this, note that there is a pronounced crossover from positive to negative MR
that occurs at high temperatures with increasing field. It is clear that the shape of the
$\rho(H)$ curves is notably different from that expected for weak-field orbital MR ($\Delta
\rho \sim H^2$), while at high fields the MR becomes negative. Both of these facts encourage
us to consider an unconventional magnetic origin to the observed MR behavior. This conclusion
is natural, taking into account the connection commonly made between zero-field $T$-linear
resistivity and AF spin fluctuations \cite{Stewart,rho_fluct}. In this scenario, the initial
increase of $\rho$ with field could originate from an increase of spin disorder. Although
positive MR is unexpected in Kondo systems \cite{H_Kondo}, an increase of MR with field is
indeed observed in systems with AF order \cite{Gegenwart,Christianson}, and in systems
approaching a coherent state \cite{HF_posMR}. In CeCoIn$_5$, there is no evidence for
long-range AF order in zero external field. However, notable AF correlations are observed below
$T_{coh}$, and therefore it is natural to associate the increase of spin disorder with a
suppression of AF correlations. Clearly, the polarization of spins by increasing field strength
should eventually lead to a field-aligned state, as was observed in the case of YbRh$_2$Si$_2$
\cite{Gegenwart}. Therefore, a crossover to negative MR should occur at progressively
increasing fields at higher temperatures, which is indeed observed in our experiment.

Based on our experimentally determined ranges of $T^2$ behavior at low temperatures, and on
the crossover observed in the sign of MR at high temperatures, we have constructed a phase
diagram of the $H-T$ plane (Fig.\ref{fig:phase}).  It is apparent that the MR crossover line
approaches the $H_{c2}$ transition at a finite temperature, so that below $\sim 1$~K the
domain of negative MR is directly adjacent to the superconducting domain. Surprisingly, this
diagram strongly resembles that determined by torque magnetometry \cite{Murphy} for $H \parallel
[110]$, where the jump in torque associated with the first-order $H_{c2}$ transition below
$\sim 1.4$~K was traced well into the normal state (up to $\sim$25~K), indicative of a
metamagnetic transformation. Although no such anomaly was found above $T_c$ for $H \parallel
[100]$ or $H\parallel [001]$, and subsequent magnetization measurements did not reproduce this
result \cite{Tayama}, the sensitivity of torque measurements may highlight the importance of
the $H \parallel [110]$ orientation. The similar behavior observed in both MR and torque
measurements further suggests the possible existence of a magnetic order parameter, where the
direct observation of a transition may be complicated by the emergence of superconductivity.

%%%%%%%%%%%%%%%%%%%%%%%%%%%% CONCLUSIONS

In conclusion, we have identified the anomalous low-temperature evolution of magnetoresistance
in CeCoIn$_5$ with the field-induced development of a Fermi liquid regime. This evolution and
the critical nature of the electron-electron scattering coefficient thus derived both bear
close resemblance to other systems governed by quantum criticality. The crossover from
positive to negative magnetoresistance, which extends to high temperatures, is indicative of a
change in character of spin fluctuations with increasing field strength, and may be closely
tied to the critical behavior observed in the resistivity, hinting at a quantum critical point
of a magnetic nature, masked by superconductivity.

%%%%%%%%%%%%%%%%%%%%%%%%%%%% ACKNOWLEDGMENTS

This work was supported by the Canadian Institute for Advanced Research and funded by NSERC. The
authors gratefully acknowledge useful discussions with J.~Hopkinson, Y.~B.~Kim, J.~L.~Sarrao, and
I.~Vekhter.

{\it Note added.-} Recent specific heat measurements by Bianchi {\it et al.} \cite{Bianchi} support
our main conclusion regarding the existence of a field-induced QCP.

%%%%%%%%%%%%%%%%%%%%%%%%%%%% BIBLIOGRAPHY


\begin{references}
\bibitem{fits} The range of $T^2$ behavior was determined using a procedure identical to that
used in Ref. \cite{Dickey}. Linear least-squares fits of the data were made over a varying
interval of temperatures, fixing the lower bound at the lowest temperature point and varying the
upper temperature bound. The reduced chi-square value was determined for each fit interval and
plotted as a function of the upper bound of the fitting range. The upper bound corresponding to
the minimum chi-square value was then chosen for the $T^2$ range.
\bibitem{upturn} This anomaly is similar to the upturn observed in the
resistivity of UPt$_3$ in transverse fields, where such a deviation from $T^2$ behavior at low
temperatures is attributed to the fact that the $\omega_c \tau > 1$ limit is realized
\cite{UPt3}. In this limit, the characteristic time of the quasiparticle orbital motion is short
compared with the time between collisions, and therefore the Fermi surface topology plays an
important role in determining transport properties. Note that, due to the presence of this
upturn, it was necessary to exclude the lowest temperature data from the $T^2$ fits (for
$H>8$~T). This was accomplished by using the same chi-square minimization routine as described
previously, but by also varying the lower temperature bound.
\end{references}
\end{document}